\documentclass[aip,pra,twocolumn,showpacs]{revtex4}
\usepackage{graphicx,color}
\usepackage{amsmath}
\usepackage{longtable}
\begin{document}

\title{Practical expressions for the internal energy and pressure of Yukawa fluids}

\author{Sergey A. Khrapak\footnote{Also at Joint Institute for High Temperatures, Russian Academy of Sciences, Moscow, Russia} and Hubertus M. Thomas }
\date{\today}
\affiliation{Forschungsgruppe Komplexe Plasmen, Deutsches Zentrum f\"{u}r Luft- und Raumfahrt,
Oberpfaffenhofen, Germany}

\begin{abstract}
Simple practical expressions are put forward, which allow to estimate thermodynamic properties of Yukawa fluids in a wide range of coupling, up to the fluid-solid phase transition.
These expressions demonstrate excellent agreement with the available results from numerical simulations. The approach provides simple and accurate tool to estimate thermodynamic properties
of Yukawa fluids and related systems in a broad range of parameters.
\end{abstract}

\pacs{05.70.Ce, 64.30.-t, 52.27.Lw}
\maketitle

\section{Introduction}

Studies of static and dynamical properties of Yukawa systems constitute an import interdisciplinary topic with applications to strongly coupled plasmas, dusty (complex) plasmas, and colloidal dispersions. An idealized model deals with point-like charged particles immersed in a neutralizing medium and interacting via the pairwise potential of Yukawa (Debye-H\"{u}ckel) type,
\begin{equation}\label{Yukawa}
V(r)= (Q^2/r)\exp(-r/\lambda_{\rm D}),
\end{equation}
where $Q$ is the particle charge, $\lambda_{\rm D}$ is the Debye screening length associated with the neutralizing medium, and $r$ is the distance between a pair of particles.
Clearly, such an idealization oversimplifies considerably the actual rather complex interactions between the particles in real systems, in particular in dusty plasmas and colloidal suspensions~\cite{IvlevBook,FortovBook,FortovRev,KhrapakPRL2008,KhrapakCPP,Morf2012}. Nevertheless, many experimentally observed trends can be reproduced by this simple consideration, at least qualitatively. Hence, it can be considered as a basis for constructing more realistic models.

Thermodynamic properties of Yukawa systems are relatively well investigated using various computational and analytical techniques. Some relevant examples include Monte Carlo (MC) and molecular dynamics (MD) numerical simulations,~\cite{Robbins,Meijer,Hamaguchi,CG} as well as integral equation theoretical studies~\cite{Tejero,Kalman2000,Faussurier,SMSA}. Semi-empirical fitting formulas~\cite{TotsujiJPA,TotsujiPoP,Vaulina2010} and simplistic approaches~\cite{DHH,ISM} to estimate thermodynamics of Yukawa systems have been also discussed in the literature. Their accuracy is in most cases not better than moderate.

The purpose of the present paper is to put forward practical approach to evaluate thermodynamic properties of Yukawa fluids across coupling regimes. This approach is based on simple phenomenological arguments, which are likely applicable to a wide class of soft repulsive interactions. In particular, expressions for the internal energy, pressure, and isothermal compressibility modulus are derived. When compared with the ``exact'' results from numerical simulations, the approach demonstrates an impressive accuracy. Hence, it represents very convenient tool to estimate thermodynamics of Yukawa and other related fluids. Among expected applications, wave-related phenomena in strongly coupled complex (dusty) plasmas can be particularly mentioned.

\section{Model}

We consider $N$ particles contained in the (three dimensional) volume $V$ and interacting via the Yukawa potential (\ref{Yukawa}). The number of particles and the volume are very large, while the number density $n_{\rm p}=N/V$ is finite, so that finite-size and surface effects are not important. The system of repelling particles is stabilized by the presence of the neutralizing medium. The effect of this neutralizing medium on the thermodynamic properties of the system can be trivially evaluated, as discussed in the Appendix~\ref{ap2}. We focus therefore on the contribution coming from particle-particle interactions and resulting correlations. Thus we effectively consider a {\it single component} Yukawa system.

The system is characterized by two dimensionless parameters. The first is the coupling parameter, $\Gamma=Q^2/aT$, where $a= (3/4\pi n_{\rm{p}})^{1/3}$ is the Wigner-Seitz radius and $T$ is the temperature (in energy units). This is roughly the ratio of the (Coulomb) interaction energy between neighboring particles to the kinetic energy. The second is the screening parameter
$\kappa=a/\lambda_{\rm D}$, which is roughly the ratio of the interparticle separation to the screening length.

The main thermodynamic quantities considered here are the internal energy $U$, Helmholtz free energy $F$, pressure $P$, and the isothermal compressibility modulus $\mu=T^{-1}(\partial P/\partial n_{\rm p})_{T}$. In reduced units these are $u=U/NT$, $f=F/NT$, and $p=PV/NT$ (the ratio $Z=PV/NT$ is also known as the compressibility factor), respectively. Except explicitly specified, we consider only the contribution from the particle-particle correlations.

\section{Derivation of energy and pressure}

The reduced excess (over that of non-interacting particles) energy can be divided into the static and thermal components
\begin{equation}\label{div}
u_{\rm ex}=u_{\rm st}+u_{\rm th}.
\end{equation}
The static contribution corresponds to the value of internal energy when the particles are frozen in some regular structure and the thermal corrections arise due the deviations of the particles from these fixed position (consequence of the thermal motion). Of course, such a division is only meaningful when the regular structure is specified. For crystals, the corresponding lattice sum is a relevant choice fort $u_{\rm st}$. For fluids, it is convenient to link $u_{\rm st}$ with the energy obtained with the Percus-Yevick (PY) radial distribution function of hard spheres in the unphysical limit $\eta =1$, where $\eta$ is the hard sphere packing fraction~\cite{Rosenfeld1998,Rosenfeld2000}. For Yukawa system this is equivalent to the result of the ion sphere model (ISM), where each particle is placed in the center of the charge neutral Wigner-Seitz spherical cell and the energy is then calculated from simple electrostatic consideration~\cite{ISM}.
With this choice, the static component of the internal energy of the single component Yukawa fluids becomes~\cite{ISM,Rosenfeld1998,Rosenfeld2000}
\begin{equation}\label{ust}
u_{\rm st}= M_{\rm f}(k)\Gamma = \frac{\kappa(\kappa+1)\Gamma}{(\kappa+1)+(\kappa-1)e^{2\kappa}},
\end{equation}
where $M_{\rm f}$ has been termed the fluid Madelung constant~\cite{Rosenfeld2000}.

\begin{figure}
\includegraphics[width=7.5cm]{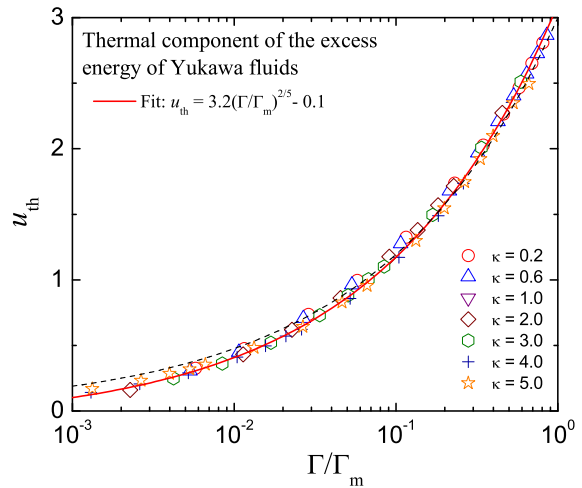}
\caption{(Color online) Thermal component of the reduced excess energy of Yukawa fluids versus the reduced coupling parameter $\Gamma/\Gamma_{\rm m}$. Symbols correspond to the numerical simulations for different values of the screening parameter $\kappa$~\cite{Farouki,Hamaguchi}. The (red) solid curve is the fit of Eq.~(\ref{fit}) with $\delta=3.2$ and $\epsilon=-0.1$. The (black) dashed line corresponds to the same functional form, but with $\delta=3.0$ and $\epsilon=0$, as suggested in~\cite{Rosenfeld2000}.}
\label{fig1}
\end{figure}

As Rosenfeld and Tarazona~\cite{Rosenfeld1998,Rosenfeld2000} first pointed out, the thermal component of the internal energy exhibits quasi-universal behavior for a wide class of soft repulsive potentials, including the Yukawa case. This is illustrated in Fig.~\ref{fig1}, where the dependence of $u_{\rm th}$ on $\Gamma/\Gamma_{\rm m}$ is plotted for a number of screening parameters $\kappa<5$.
Here $\Gamma_{\rm m}$ is the value of the coupling parameter at the solid-fluid phase transition (melting). To produce this plot, the numerical data on $u_{\rm ex}(\kappa,\Gamma)$ and $\Gamma_{\rm m}(\kappa)$ tabulated in Refs.~\cite{Hamaguchi,Farouki} have been used (the contribution of the neutralizing medium has been subtracted). It is seen that numerical data have a tendency to group around a single quasi-universal curve. Reasonably accurate fits (shown by the curves) can be obtained by using the following functional form
\begin{equation}\label{fit}
u_{\rm th}=\delta (\Gamma/\Gamma_{\rm m})^{2/5}+\epsilon.
\end{equation}
The original suggestion of Rosenfeld to use $\delta=3.0$ and $\epsilon=0$~\cite{Rosenfeld2000} is shown by the dashed curve. Some improvement can be observed when choosing $\delta=3.2$ and $\epsilon=-0.1$, as documented in Figure~\ref{fig1}. These values are therefore adopted throughout this paper. Note that although the functional form (\ref{fit}) with the exponent $\tfrac{2}{5}$ provides reasonable accuracy for Yukawa fluids in a wide range of $\kappa$, it can be not the best choice for each single value of $\kappa$. For instance, in the case of one-component-plasma (limiting case $\kappa=0$ of Yukawa systems), the exponent $\tfrac{1}{3}$ is known to deliver better accuracy~\cite{Farouki,Stringfellow,Dubin,KK2014}.

Equations (\ref{div})-(\ref{fit}) with the proper expression for $\Gamma_{\rm m}(\kappa)$ (see below) provide a simple and accurate tool to estimate the excess energy of Yukawa fluids. The excess free energy can then be obtained by the standard integration
\begin{equation}\label{free}
f_{\rm ex}(\kappa,\Gamma)= \int_0^{\Gamma}\frac{u_{\rm ex}(\kappa,\Gamma')}{\Gamma'}d\Gamma'.
\end{equation}
Note that in case of the non-zero value of the parameter $b$, this integral is diverging logarithmically. A simple conventional procedure to avoid this divergence is to start integration from $\Gamma=1$ in (\ref{free}) and add the corresponding value $f_{\rm ex}(\kappa,1)$~\cite{Hamaguchi}. The contribution from the weakly coupled region $f_{\rm ex}(\kappa,1)$ is in fact of minor importance for $\Gamma\gg 1$. The values of $f_{\rm ex}(\kappa,1)$ have been tabulated in Refs.~\cite{Hamaguchi,Farouki}, interpolation is straightforward. This is not done here, because the main interest is the excess pressure and its derivative at strong coupling, where the contribution from the weak coupling region is negligible.

Using the dimensionless quantities $\kappa$ and $\Gamma$ instead of particle density and temperature, the excess pressure can be obtained from~\cite{SMSA,DHH}
\begin{equation}\label{p1}
p_{\rm ex}(\kappa,\Gamma)= \frac{\Gamma}{3}\frac{\partial f_{\rm ex} }{\partial \Gamma}-\frac{\kappa}{3}\frac{\partial f_{\rm ex}}{\partial \kappa}.
\end{equation}
After the straightforward manipulation we can rewrite this as
\begin{equation}\label{pex}
p_{\rm ex}(\kappa,\Gamma)=p_{\rm st}(\kappa,\Gamma)+\frac{1}{3}u_{\rm th}(\kappa,\Gamma)-\frac{5\kappa}{6}\frac{\partial u_{\rm th}(\kappa, \Gamma)}{\partial \kappa},
\end{equation}
where $p_{\rm st}(\kappa,\Gamma)=\tfrac{1}{3}\Gamma[M_{\rm f}(\kappa)-\kappa\partial M_{\rm f}(\kappa)/\partial \kappa]$ is the excess pressure associated with the static component of the internal energy and the two last terms in (\ref{pex}) come from the thermal component of the internal energy. The static contribution can be easily evaluated and yields a compact expression~\cite{ISM}
\begin{equation}\label{pst}
p_{\rm st}(\kappa,\Gamma)= \frac{\kappa^4\Gamma}{6\left[\kappa\cosh(\kappa)-\sinh(\kappa)\right]^2}.
\end{equation}
To evaluate the thermal contribution, the functional dependence $\Gamma_{\rm m}(\kappa)$ should be specified. Several approximate methods to locate the fluid-solid phase transition, based on the properties of the interaction potential alone, have been discussed in the literature~\cite{Rosenfeld1976,KhrapakPRL,Univers,KhSa}. Some of them are applicable to a wide range of interactions, including the Yukawa case, and can be used for this purpose. Being focused on Yukawa interaction here, we adopt a simple expression proposed by Vaulina {\it et al}.~\cite{VaulinaJETP,VaulinaPRE}
\begin{equation}\label{melt}
\Gamma_{\rm m}(\kappa)\simeq \frac{172 \exp(\alpha\kappa)}{1+\alpha\kappa+\tfrac{1}{2}\alpha^2\kappa^2},
\end{equation}
where the constant $\alpha=(4\pi/3)^{1/3}\simeq 1.612$  is the ratio of the mean interparticle distance $\Delta=n_{\rm p}^{-1/3}$ to the Wigner-Seitz radius $a$. Equation (\ref{melt}) is in rather good agreement with the numerical data in the regime $\kappa\lesssim 5$ addressed in this study. Equations (\ref{fit}), (\ref{pex}), (\ref{pst}), and (\ref{melt}) allow to evaluate the excess pressure and hence the compressibility factor,
\begin{equation}
Z(\kappa,\Gamma)=1+p_{\rm ex}(\kappa,\Gamma),
\end{equation}
of Yukawa fluids for any given pair $\kappa$ and $\Gamma$. This constitutes the main result of this paper. In the Appendix~\ref{ap1} the explicit expressions for the compressibility factor and isothermal compressibility modulus are provided, which are convenient for practical calculations.

\section{Comparison with previous studies}

\begin{table}
\caption{\label{Tab0} Compressibility factor (reduced pressure)  of a single component Yukawa fluid in a wide range of coupling. The first two columns specify the location of the system state point on the $(\kappa,\Gamma)$ plane. The third column lists the values of the reduced coupling strength $\Gamma/\Gamma_{\rm m}$ (note that the first point may correspond to supercooled liquid). The remaining columns contain the values of $Z$ obtained using MC simulations by Meijer and Frenkel (MF)~\cite{Meijer} ($Z_{\rm MF}$), DRY method by Tejero {\it et al}.~\cite{Tejero} ($Z_{\rm DRY}$), present approach ($Z_{\rm present}$), and its static component ($Z_{\rm st}$). For details see the text.}
\begin{ruledtabular}
\begin{tabular}{lllllll}
$\kappa$ & $\Gamma$ & $\Gamma/\Gamma_{\rm m}$ &  $Z_{\rm MF}$ & $Z_{\rm DRY}$ & $Z_{\rm present}$ & $Z_{\rm st}$  \\ \hline
1.800 & 396.9 & 1.03 & 102.492 & 102.751 &  102.526  & 99.856\\
1.860 & 383.9 & 0.95 & 89.606 & 89.846 & 89.567   & 86.902 \\
1.923 & 371.4 & 0.87 & 78.148 & 78.387 & 78.145  & 75.487 \\
1.984 & 360.0 & 0.80 & 68.640 & 68.865 & 68.637 & 65.987 \\
2.049 & 348.6 & 0.73 & 59.889 & 60.091 & 59.895 & 57.256 \\
2.117 & 337.5 & 0.66 & 52.133 & 52.307 & 52.150 & 49.523 \\
2.182 & 327.3 & 0.60 & 45.711 & 45.862 & 45.707 & 43.095 \\
2.238 & 319.2 & 0.56 & 41.041 & 41.176 & 41.002 & 38.404 \\
2.306 & 309.7 & 0.51 & 35.954 & 36.072 & 35.903 & 33.324 \\
2.348 & 304.2 & 0.48 & 33.204 & 33.314 & 33.184 & 30.618 \\
2.398 & 297.9 & 0.45 & 30.294 & 30.394 & 30.249 & 27.699 \\
2.532 & 282.1 & 0.37 & 23.780 & 23.855 & 23.741 & 21.237 \\
2.631 & 271.5 & 0.32 & 20.016 & 20.069 & 19.989 & 17.524 \\
2.778 & 257.1 & 0.26 & 15.705 & 15.722 & 15.682 & 13.279 \\
3.050 & 234.2 & 0.18 & 10.400 & 10.343 & 10.418 & 8.144 \\
\end{tabular}
\end{ruledtabular}
\end{table}

To demonstrate the accuracy of the present approach, the compressibility factor of a single component Yukawa fluid has been evaluated in a wide range of coupling strength and compared with the results available in the literature. This comparison is shown in Table~\ref{Tab0}. The first three columns contain the location of the system state point in terms of $\kappa$, $\Gamma$, and $\Gamma/\Gamma_{\rm m}$, respectively. The fourth column lists the results from the MC simulations performed by Meijer and Frenkel (MF)~\cite{Meijer} and tabulated in Ref.~\cite{Tejero}, which serve as a reference data here. The fifth column contains the results obtained using the modified Rogers-Young (RY) integral equation~\cite{Tejero}. This method requires less numerical efforts compared to the original RY implementation, but the thermodynamic consistency is achieved only on a discrete grid of points and for chosen fitting functions~\cite{Tejero}. It is therefore referred to as the discretized RY method (DRY). The sixth column summarizes the results obtained using the present approach, Eq.~(\ref{expl_p}). The last column provides the static contribution to the compressibility factor, $Z_{\rm st}(\kappa,\Gamma)=1+p_{\rm st}(\kappa,\Gamma)$.

The simple approximation discussed in the present paper is in excellent agreement with the precise results from MC simulations by Meijer and Frenkel. The deviations from their numerical data do not exceed few parts in one thousand in all cases considered and are significantly smaller than that at strong coupling, near the fluid-solid phase transition. Present approach demonstrates better accuracy than the DRY integral equation method. The static component of the compressibility provides reasonable estimate of the actual compressibility near freezing ($\sim 3\%$ underestimation), but becomes progressively less accurate when coupling decreases.

Since the numerical data tabulated in Ref.~\cite{Tejero} are limited to a relatively narrow range of the screening parameter $\kappa$, we have performed additional comparison with the MC simulations of Yukawa systems on a hypersphere, performed by Caillol and Gilles (CG)~\cite{CG}. The highest value of the coupling parameter $\Gamma=100$ investigated in Ref.~\cite{CG} has been chosen for detailed comparison. The values of the compressibility factor obtained in MC simulation and those calculated with the help of the expression (\ref{fit}) are listed in Table~\ref{Tab1} for a number of $\kappa$-values ($0.1\leq\kappa\leq 5.0$). To keep the original notation of Ref.~\cite{CG} we have added the contribution of the neutralizing medium to the compressibility factor (see Appendix~\ref{ap2}), that is why the compressibility is negative. The agreement is again excellent at sufficiently strong coupling. The relative deviations do not exceed a tiny fraction of a percent as long as $\kappa\lesssim 3.0$, where $\Gamma/\Gamma_{\rm m}$ has dropped well below 0.1. For higher $\kappa$, deviations increase and  reach several percent. This apparently indicates that the contribution from the weak coupling region has to be properly accounted for in this regime. Overall, the present approach demonstrates excellent performance at least in the regime $\kappa\lesssim 5$ and $\Gamma/\Gamma_{\rm m}\gtrsim 0.1$ and therefore can find application in many practical situations relevant to colloidal systems and complex plasmas.

\begin{table}
\caption{\label{Tab1} Compressibility factor (reduced pressure) of the conventional Yukawa fluid (with neutralizing background) for a fixed coupling parameter $\Gamma= 100$ and various screening parameters $\kappa$. The first two columns specify the location of the system state point on the $(\kappa,\Gamma/\Gamma_{\rm m})$ plane. The third and fourth columns contain the values of $\tilde{Z}$ obtained using MC simulations by Caillol and Gilles (CG)~\cite{CG} and using the present approach (present), respectively.
%The last column displays absolute values of relative deviation (RD) in percents.
}
\begin{ruledtabular}
\begin{tabular}{lllll}
$\kappa$ & $\Gamma/\Gamma_{\rm m}$ & $\tilde{Z}_{\rm CG}$ &  $\tilde{Z}_{\rm present}$ \\ \hline
0.1 &  0.58  & -28.1399  & -28.1391  \\
0.2 &  0.58  & -28.0233  & -28.0291  \\
0.4 &  0.57  & -27.5706  & -27.5819  \\
0.6 &  0.54  & -26.8387  & -26.8495  \\
0.8 &  0.50  & -25.8645  & -25.8698  \\
1.0 &  0.45  & -24.6883  & -24.6885  \\
1.4 &  0.35  & -21.9057  & -21.9122  \\
2.0 &  0.22  & -17.3432  & -17.3674  \\
2.5 &  0.14  & -13.7956  & -13.7996  \\
3.0 &  0.08  & -10.8036  & -10.7379  \\
3.5 &  0.05  & -8.43152  & -8.26628  \\
4.0 &  0.03  & -6.60858  & -6.34905  \\
5.0 &  0.01  & -4.15975  & -3.80722  \\

\end{tabular}
\end{ruledtabular}
\end{table}

\section{Conclusion}

Simple approach to estimate the internal energy, pressure, and compressibility modulus of three-dimensional Yukawa fluids across coupling regimes has been put forward. Explicit analytical expressions for these quantities have been derived (see Appendix~\ref{ap1}), which demonstrate excellent agreement with precise results from MC simulations at strong coupling. These expressions are directly applicable to single component Yukawa systems, modifications to describe charged particles in the neutralizing medium are trivial (see Appendix~\ref{ap2}). The obtained results can be particularly useful in connection with studying wave phenomena in strongly coupled complex (dusty) plasmas, since simple and accurate equation of state is often required for such studies. For this reason, the main focus here has been on pressure-related quantities, although other thermodynamic functions can apparently be also estimated using the proposed approach. The approach is also likely to be relevant for other simple fluids with soft repulsive interactions, when the static component of the internal energy dominates over the thermal one. This and related issues are left for future work.

\begin{acknowledgments}
We would like to thank Panagiotis Tolias for useful discussions. This study was partially supported by the Russian Science Foundation, Project No. 14-12-01235.
\end{acknowledgments}

\appendix

\begin{widetext}

\section{Practical expressions for compressibility factor and isothermal compressibility modulus}\label{ap1}

To keep some generality we consider the coefficients $\delta$ and $\epsilon$, entering Eq. (\ref{fit}) for the thermal component of the excess energy, as unspecified for a moment.
The explicit expression for the compressibility factor is
\begin{equation}\label{expl_p}
Z(\kappa,\Gamma)=\left(1+\frac{\epsilon}{3}\right)+ \frac{\Gamma\kappa^4}{6\left[\kappa\cosh(\kappa)-\sinh(\kappa)\right]^2}+\frac{\delta}{3}\left(\frac{\Gamma}{\Gamma_{\rm m}}\right)^{2/5}f_{\rm Z}(\alpha\kappa),
\end{equation}
where
\begin{equation}
f_{\rm Z}(x)=\frac{x^3+x^2+2x+2}{x^2+2x+2}.
\end{equation}
The isothermal compressibility modulus is related to the compressibility factor via $\mu=Z+(\Gamma/3)(\partial Z/\partial \Gamma)-(\kappa/3)(\partial Z/\partial \kappa)$. This yields
\begin{equation}\label{expl_mu}
\mu(\kappa,\Gamma)=\left(1+\frac{\epsilon}{3}\right)+ \frac{\Gamma\kappa^6\sinh(\kappa)}{9\left[\kappa\cosh(\kappa)-\sinh(\kappa)\right]^3}+\frac{\delta}{45}\left(\frac{\Gamma}{\Gamma_{\rm m}}\right)^{2/5}f_{\rm \mu}(\alpha\kappa),
\end{equation}
where
\begin{equation}
f_{\rm \mu}(x)=\frac{2x^6+14x^5+35x^4+76x^3+136x^2+136x+68}{(x^2+2x+2)^2}.
\end{equation}
The dependence $\Gamma_{\rm m}(\kappa)$ is given by Eq.~(\ref{melt}). Regarding the coefficients $\delta$ and $\epsilon$, we suggest to use $\delta=3.2$ and $\epsilon=-0.1$.

\end{widetext}

\section{Effect of the neutralizing medium}\label{ap2}

The contributions from particle-particle correlations and particle-background interactions are additive. In particular, the excess energy associated with the presence of neutralizing medium is~\cite{Ham94}
\begin{equation}
u_{\rm m}=-\frac{3\Gamma}{2\kappa^2}-\frac{\kappa\Gamma}{2}.
\end{equation}
The first term represents the excess (free) energy of the medium that, on average, neutralizes the
charge of the particles while the second term gives the (free) energy of the sheath around each particle. Note that this latter term does not contribute to the excess pressure, as can be clearly seen from Eq.~(\ref{p1}). It, therefore, has no effect on the compressibility modulus, too. To account for the effect of the neutralizing medium, one simply needs to add the term $-3\Gamma/2\kappa^2$ to the compressibility factor, Eq.~(\ref{expl_p}), and the term $-3\Gamma/\kappa^2$ to the isothermal compressibility modulus, Eq.~(\ref{expl_mu}). Note that the contribution of the neutralizing medium is {\it negative} and {\it dominant} at strong coupling, implying that the excess energy, pressure, and compressibility are also negative in this regime.


\begin{thebibliography}{99}

\bibitem{IvlevBook} A. Ivlev, H. L\"{o}wen, G. Morfill, and C. P. Royall, {\it Complex Plasmas and Colloidal Dispersions: Particle-resolved Studies of Classical Liquids and Solids} (World Scientific, Singapore, 2012).
\bibitem{FortovBook} {\it Complex and dusty plasmas: From Laboratory to Space}, edited by V. E. Fortov and G. E. Morfill (CRC Press, Boca Raton, 2010).

\bibitem{FortovRev} V. E. Fortov, A. G. Khrapak, S. A. Khrapak, V. I. Molotkov, and O. F. Petrov,
Phys. Usp. {\bf 47}, 447 (2004); Fortov, A. V. Ivlev, S. A. Khrapak, A. G. Khrapak, and G. E. Morfill, Phys. Rep. {\bf 421}, 1 (2005).
\bibitem{KhrapakPRL2008} S. A. Khrapak, B. A. Klumov, G. E. Morfill, Phys. Rev. Lett. {\bf 100}, 225003 (2008).
\bibitem{KhrapakCPP} S. Khrapak and G. Morfill, Contrib. Plasma Phys. {\bf 49}, 148 (2009).
\bibitem{Morf2012} G. E. Morfill, A. V. Ivlev, and H. M. Thomas, Phys. Plasmas {\bf 19}, 055402 (2012).

\bibitem{Robbins} M. O. Robbins, K. Kremer, and G. S. Grest, J. Chem. Phys. {\bf 88}, 3286 (1988).
\bibitem{Meijer} E. J. Meijer and D. Frenkel, J. Chem. Phys. {\bf 94}, 2269 (1991).
\bibitem{Hamaguchi} S. Hamaguchi, R. T. Farouki, and D. H. E. Dubin, Phys. Rev. E {\bf 56}, 4671 (1997).
\bibitem{CG} J. M. Caillol and D. Gilles, J. Stat. Phys. {\bf 100}, 933 (2000).

\bibitem{Tejero} C. F. Tejero, J. F. Lutsko, J. L. Colot, and M. Baus, Phys. Rev. A {\bf 46}, 3373 (1992).
\bibitem{Kalman2000} G. J. Kalman, M. Rosenberg, and H. DeWitt, J. Phys. IV France {\bf 10}, 403 (2000).
\bibitem{Faussurier} G. Faussurier, Phys. Rev. E {\bf 69}, 066402 (2004).
\bibitem{SMSA} P. Tolias, S. Ratynskaia, and U. de Angelis, Phys. Rev. E {\bf 90}, 053101 (2014).

\bibitem{TotsujiJPA} H. Totsuji, J. Phys. A: Math. Gen. {\bf 39}, 4565 (2006).
\bibitem{TotsujiPoP} H. Totsuji, Phys. Plasmas {\bf 15}, 072111 (2008).
\bibitem{Vaulina2010} O. S. Vaulina, X. G. Koss, Yu. V. Khrustalev, O. F. Petrov, and V. E. Fortov, Phys. Rev. E {\bf 82}, 056411 (2010).

\bibitem{DHH} S. A. Khrapak, A. G. Khrapak, A. V. Ivlev, and G. E. Morfill, Phys. Rev. E {\bf 89}, 023102 (2014).
\bibitem{ISM} S. A. Khrapak, A. G. Khrapak, A. V. Ivlev, and H. M. Thomas, Phys. Plasmas (2014, in press).

\bibitem{Rosenfeld1998} Y. Rosenfeld and P. Tarazona, Mol. Phys. {\bf 95}, 141 (1998).
\bibitem{Rosenfeld2000} Y. Rosenfeld, Phys. Rev. E {\bf 62}, 7524 (2000).


\bibitem{Farouki} R. T. Farouki and S. Hamaguchi, J. Chem. Phys. {\bf 101}, 9885 (1994).

\bibitem{Stringfellow} G. S. Stringfellow, H. E. DeWitt, and W. L. Slattery, Phys. Rev. A {\bf 41}, 1105 (1990).
\bibitem{Dubin} D. H. E. Dubin and T. M. O'Neil, Rev. Mod. Phys. {\bf 71}, 87 (1999).
\bibitem{KK2014} S. A. Khrapak and A. G. Khrapak, Phys. Plasmas {\bf 21}, 104505 (2014).

\bibitem{Rosenfeld1976} Y. Rosenfeld, J. Chem. Phys. {\bf 64}, 1248 (1976); Mol. Phys. {\bf 32}, 963 (1976).
\bibitem{KhrapakPRL} S. A. Khrapak and G. E. Morfill, Phys. Rev. Lett. {\bf 103}, 255003 (2009).
\bibitem{Univers} S. A. Khrapak, M. Chaudhuri, and G. E. Morfill, J. Chem. Phys. {\bf 134}, 241101 (2011).
\bibitem{KhSa} S. A. Khrapak and F. Saija, Mol. Phys. {\bf 109}, 2417 (2011).

\bibitem{VaulinaJETP} O. S. Vaulina and S. A. Khrapak, JETP {\bf 90}, 287 (2000).
\bibitem{VaulinaPRE} O. Vaulina, S. Khrapak, and G. Morfill, Phys. Rev. E {\bf 66}, 016404 (2002).



\bibitem{Ham94} S. Hamaguchi and R. T. Farouki, J. Chem. Phys. {\bf 101}, 9876 (1994).

\end{thebibliography}
\end{document}